\documentstyle[11pt,moriond,epsfig]{article}
\def\Journal#1#2#3#4{{#1} {\bf #2}, #3 (#4)}

\def\NPB{{\em Nucl. Phys.} B}
\def\PLB{{\em Phys. Lett.}  B}
\def\PRL{\em Phys. Rev. Lett.}
\def\PRD{{\em Phys. Rev.} D}

\begin{document}
\vspace*{4cm}
\title{POST-CLASSICISM IN TREE AMPLITUDES}

\author{ K.G. SELIVANOV }

\address{ITEP, B.Cheremushkinskaya,25, Moscow, Russia}

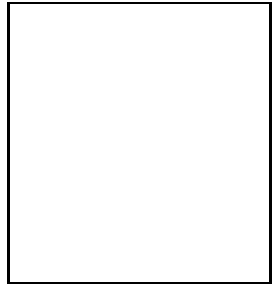
\maketitle\abstracts{
This is a short and simple introduction into {\it perturbiners},
that is, the solutions of field equations which are generating 
functions for tree amplitudes. The perturbiners have been constructed
in Yang-Mills, in SUSY Yang-Mills, in gravity, and in sin(h)-Gordon.}

\newpage

I would like to present here work done in collaboration with Alexei Rosly.
This is devoted to solutions of field equations which describe tree 
amplitudes with any number of external legs. In works \cite{KS1}, \cite{RS1},
\cite{KS2} such solutions were constructed in Yang-Mills (see also
a closely related work \cite{Ba}). In works \cite{RS2}, \cite{KS3},
\cite{KS4} such solutions were constructed in gravity and in Yang-Mills 
interacting with gravity. Work \cite{RS3} is about the sin(h)-Gordon case
and work \cite{KS5} deals with the case of SUSY Yang-Mills.  

Unfortunately, because of lack of the time and because of the interest
of the audience, I shall not be able to explain the constructions itself -
you will have to trust me that they are beautiful. I shall only explain
what type of solutions we construct. So to say, I shall only formulate the 
problem, a solution for which can be found in the references above.
 
First of all, I, perhaps, need to explain that multi-leg amplitudes are 
non-trivial objects even in the tree approximation. The problem is that, 
although every given diagram gives quite a simple contribution, an algebraic 
function of external momenta, the number of different diagrams growths
enormously with the number of legs, and all these algebraic functions are 
different, so that the whole thing becomes untreatable when the number of 
legs becomes bigger than, say, 10, to say nothing about arbitrary number of 
legs. 

In principle, one could take this as that the nature is such that
the multi-leg amplitudes are complicated objects and nothing to do
about it. However, there are known cases when the final expression for the 
amplitudes  - if it is available at all - turns out  to be much simpler than
intermediate steps. For instance, in \cite{V}, \cite{S}, \cite{A} it was 
noticed that in ${\phi}^{4}$ theory all amplitudes $2 {\rightarrow} n$
for $n$ bigger than 4 vanish at threshold. Notice that every given diagram 
gives a nontrivial contribution. That is only sum of them  what vanishes.

Other example of the case when the final expression is much simpler than
intermediate steps are the so-called Parke-Taylor, or ``maximally helicity
violating'' amplitudes in Yang-Mills \cite{PT}, \cite{BG}. These are 
amplitudes with two positive-helicity gluons in the initial state and any 
number of the positive-helicity gluons in the final state. Kinematics is
arbitrary. The explicit expressions for these amplitudes were conjectured 
in \cite{PT} and were proven in \cite{BG}. I am not giving here those nice
expressions, I just notice that they contain only the pairwise collinear
singularities, those of the type of $\frac{1}{(p_{i},p_{j})}$, where
$p_{i}$ stands for momentum of $i$-th particle, while a separate diagram
definitely gives much more singularities.

These remarkable cancellations above motivated us for the work presented here.

Our initial point is that tree amplitudes are related to a solution of the 
classical field equation of the model. To make the relation precise we need 
to consider not the amplitudes themself, but form-factors, that is,
would be amplitudes with one unamputated off-shell leg and a number of
on-shell ones, which are amputated as they should be according to the LSZ
rules. It is convenient to take the off-shell leg in the coordinate $(x)$
representation. Of course, the amplitudes can be obtained from the 
form-factors applying the LSZ rules to the off-shell leg.

The form-factors can also be written as matrix elements of the field operator
between vacuum and $n$-particle state, 
\begin{equation}
\label{ff}
<p_{1}, \ldots , p_{l}|{\phi}(x)|0>_{tree}
\end{equation}
where the subscript indicates that we are interested only in the tree
contributions. At tree level, when analytical structure of the form-factor
is very simple (it is just an algebraic function of the momenta), we can 
afford not to make a difference between in- and out- states. 

Notice that $x$-dependence of the form-factors is very simple, it is just a 
product of the plane waves corresponding to the momenta of external on-shell
legs. This follows from the fact that in momentum representation, the 
dependence on the momentum of the off-shell leg is given by the overall
conservation-law ${\delta}$-function:
\begin{equation}
\label{harmonics}
<p_{1}, \ldots , p_{l}|{\phi}(x)|0>_{tree}=
{\int}dp' e^{ip'x} {\delta}(p'-{\sum}_{j}p_{j})( \ldots )=
e^{i{\sum}_{j}p_{j}x}(\;a\;function\;of\;\{p_{j}\})
\end{equation}

Obviously, an individual form-factor does not obey field equations. What 
obeys field equations is a generating function for form-factors which we 
define next. To do this we\\ 
1)fix a set of momenta $\{p_{j}\}: \; p_{1}, \ldots , p_{N}$;\\
2)introduce corresponding set of parameters 
$\{a_{j}\}:\; a_{1}, \ldots , a_{N}$;\\
3)define a function of $x$, $\{p_{j}\}$ and $\{a_{j}\}$ such that the 
individual form-factors appears as coefficients in the Taylor expansion 
of this function in powers of $a$'s:
\begin{equation}
\label{ptb}
{\Phi}(x, \{p\}, \{a\})=\sum_{l=1}^{L}\sum_{\{J\}}a_{J_{1}} \ldots 
a_{J_{l}}<p_{J_{1}}, \ldots , p_{J_{l}}|{\phi}(x)|0>_{tree}
\end{equation}
where the one-particle form-factor is clearly
\begin{equation}
\label{norm}
<p|{\phi}(x)|0>=( \ldots ) e^{ipx}
\end{equation}
$( \ldots )$ in Eq.(\ref{norm}) stands for a polarization factors in a
non-scalar case, as well as for a color matrix in the Yang-Mills case, etc.

So defined generating function can be seen to obey field equations of the 
model. Indeed, applying the inverse propagator to the off-shell leg of an 
individual form-factor one obtains a sum of products of two, of three
and so on - corresponding to the types of vertices in the theory - 
form-factors with smaller amount of legs. This relation can be used as a 
recursion relation. On the other hand, this relation is seen to be equivalent
to field equations on the generating function:
\begin{equation}
\label{eq}
\left( \frac{{\partial}^{2}}{{\partial}x^{2}} + m^{2} \right)
{\Phi}(x,\{p\},\{a\})={\lambda}_{3}{\Phi}^{2}+{\lambda}_{4}{\Phi}^{3}+ \ldots
\end{equation}
Indeed, expanding Eq.(\ref{eq}) in powers of $a$'s one obtains the recursion
relation among form-factors. (Eq.(\ref{eq}) is written for a scalar theory
just for the brevity.)

Having started with the form-factors we have proved that the generating
function obeys field equations. We can reverse the logic and use the field 
equations to find ${\Phi}(x,\{p\},\{a\})$. To do this we must specify what
solution to pick up, and according to the definition above we should 
pick up a solution which is a power series in the set of variables
\begin{equation}
\label{harmonics2}
{\cal E}_{j}=a_{j}e^{ip_{j}x}, \; j=1, \ldots, N
\end{equation}
starting with first order terms of the type of
\begin{equation}
\label{expansion}
{\Phi}(x,\{p\},\{a\})={\sum}_{j=1}^{N}( \ldots){\cal E}_{j} + \;
higher\; order\; terms\; in\; \{{\cal E}_{j}\}
\end{equation}
$( \ldots)$ here is the same as in Eq.(\ref{norm}).

Clearly, solution of this type  exists and is unique (in gauge theories
it is unique after a gauge fixing, or, equivalently, it is unique modulo
gauge transformations), provided the operator on r.h.s. of Eq.(\ref{eq})
is invertible on the power series in the variables ${\cal E}_{j}$, which 
is true when the {\it nonresonantness condition} is satisfied, that is,
none of linear combinations with integer coefficients of the momenta
$p_{j}$ from the given set gets to mass shell, 
\begin{equation}
\label{nr}
({{\sum}_{j}n_{j}p_{j}})^{2}{\neq}m^{2}
\end{equation}
This is  simply a condition 
that there are no  internal lines on-shell in the Feynman diagrams.

It is also convenient to impose the {\it nilpotency} condition on the 
parameters $a_{j}$,
\begin{equation}
\label{nil}
a_{j}^{2}=0, \; j=1, \ldots, N
\end{equation}
(note that in bosonic case still $a_{i}a_{j}=a_{j}a_{i}$)
which is equivalent to excluding form-factors with identical particles
from the generating function. In the massive case this condition is 
very convenient technically, while in the massless case it is even
problematic to proceed without the nilpotency, because in the massless case
identical particles necessary break nonresonantness.

I would like to stress at this point that we have introduced a class of
solutions of field equations which are as universal as, say, solitonic
solutions. We called these solutions {\it perturbiners}.

In the classical text-books on quantum field theory,
e.g. \cite{FS},\cite{IZ}, one can also find a discussion of solutions
generating tree amplitudes, but they are defined differently, with use
of asymptotic Feynman-type conditions. That definition is not very
convenient, and it is not strange that no explicit examples were given in 
those books.

In the papers cited at the beginning of this talk we constructed 
perturbiners in various theories. Unfortunately, in Yang-Mills and in 
gravity the generic perturbiner is unavailable so far. What allows one to
proceed is the restriction on polarizations of the external on-shell
particles in the form-factors. Restricted in this way perturbiner obeys
not the full Yang-Mill (Einstein) equations, but just the self-duality 
equations instead. It generates just the same-helicity form-factors.
Having obtained the specific perturbiner with only
same-helicity form-factors included, one can add opposite helicity
particles perturbatively. This has been done in \cite{RS1} in order
to describe the Parke-Taylor amplitudes and in \cite{KS4} to generalize
the Parke-Taylor amplitudes to include any number of the same helicity
gravitons in addition to the gluons. Our main tool was the so-called
zero-curvature representation of the field equations. For the self-duality
equations (in Yang-Mills and in gravity) and for the sin(h)-Gordon
the zero-curvature representation is one-dimensional and this the case
when perturbiner is constructed very efficiently. Remarkably, 
same-helicity form-factors in Yang-Mills and generic (tree) form-factors
in sin(h)-Gordon are described in the same way. Perhaps, this opens a 
perspective to find Yang-Mills avatars of the many known quantum exact 
results in sin(h)-Gordon. In \cite{KS5} a progress
has been achieved about generic perturbiner in $N=3$ SUSY Yang-Mills
(which, of course, contains complete information about all tree
form-factors of non-supersymmetric Yang-Mills). However, the 
zero-curvature representation is two-dimensional in that case,
the construction is much more involved, and the complete solution has
not been obtained yet.  

In conclusion, I would like to stress that perturbiners provide
probably the simplest way of describing tree amplitudes, involve
nice mathematical constructions and reveal similar structures underlying 
different theories. I hope they will find their place in future
developments.

\section*{Acknowledgments}
I would like to acknowledge Russian Ministry of Sciences and
organizers of the conference for a financial support.


\section*{References}
\begin{thebibliography}{99}
\bibitem{KS1} K.G.Selivanov,  ITEP-21-96,  hep-ph/9604206
\bibitem{RS1} A.A.Rosly and K.G.Selivanov, \Journal{\PLB}{399}{135}{1997}
\bibitem{KS2} K.G.Selivanov, Talk given at International Europhysics 
Conference on High-Energy Physics (HEP 97), Jerusalem, Israel, 19-26 Aug 1997. 
\bibitem{Ba} W.Bardeen, {\em Prog.Theor.Phys.Suppl.} 123 (1996) 1
\bibitem{RS2} A.A.Rosly and K.G.Selivanov,  ITEP-TH-56-97,  hep-th/9710196
\bibitem{KS3} K.G.Selivanov, \Journal{\PLB}{420}{274}{1998}
\bibitem{KS4} K.G.Selivanov, {\em Mod.Phys.Lett.A} 12 {1997} 3087
\bibitem{RS3}  A.A.Rosly and K.G.Selivanov, \Journal{\PLB}{426}{334}{1998}
\bibitem{KS5} K.G.Selivanov,  ITEP-TH-47-98,  hep-th/9809046
\bibitem{V} M.B.Voloshin, \Journal{\PRD}{47}{357}{1993}
\bibitem{S} B.Smith, \Journal{\PRD}{47}{3518}{1993}
\bibitem{A} E.Argyres, R.Kleiss and C.Papadopoulos, \Journal{\PLB}
{302}{70}{1993}
\bibitem{PT} S.Parke and T.Taylor, \Journal{\PRL}{56}{2459}{1986} 
\bibitem{BG} F.Berends and W.Giele, \Journal{\NPB}{306}{759}{1988}
\bibitem{FS} L.D.Faddeev and A.A.Slavnov, Introduction to the Theory\\ 
of Quantum Gauge Fields, Nauka, Moscow, 1978        
\bibitem{IZ} C.Itzykson and J.B.Zuber, Quantum Field Theory\\
 New York, Usa: Mcgraw-hill (1980)  
\end{thebibliography}
\end{document}